# Calculation of Critical Nucleation Rates by the Persistent Embryo Method: Application to Quasi Hard Sphere Models

Shang Ren[1,2], Yang Sun[2*], Feng Zhang[2*], Alex Travesset[1,2], Cai-Zhuang Wang[1,2], and Kai-Ming Ho[1,2,3]

[1]*Department of Physics and Astronomy, Iowa State University, Ames, Iowa 50011, USA*

[2] *Ames Laboratory, U.S. Department of Energy, Ames, Iowa 50011, USA*

[3]*Hefei National Laboratory for Physical Sciences at the Microscale and Department of Physics, University of Science and Technology of China, Hefei, Anhui 230026, China*

**Abstract**

We study crystal nucleation of the Weeks-Chandler-Andersen (WCA) model, using the recently introduced Persistent Embryo Method (PEM). The method provides detailed characterization of pre-critical, critical and post-critical nuclei, as well as nucleation rates that compare favorably with those obtained using other methods (umbrella sampling, forward flux sampling or seeding). We further map our results to a hard sphere model allowing to compare with other existing predictions. Implications for experiments are also discussed.

## 1. Introduction

The nucleation of a crystal out of a supercooled fluid is a fundamental process of enormous complexity. Even though Classical Nucleation Theory (CNT) was proposed almost a century ago[1], it is only recently, with the development of powerful computers, that parameter-free predictions have become possible. Hard sphere (HS) models[2,3] and other closely related systems[4–8] have provided fertile test grounds on which implementations of CNT can be tested: Brute force Brownian dynamics (BD)[4,5], forward flux sampling (FFS)[2,4,6], umbrella sampling (US)[4], metadynamics and seeding method[7]. Despite its obvious theoretical motivations, these models are also of considerable experimental interest, as they approximately describe colloidal dispersions[9–

---

[*]Email: yangsun@ameslab.gov (Y.S.) or fzhang@ameslab.gov (F.Z.)

[11]. Experimental nucleation rates, however, remain in significant disagreement with most existing theoretical predictions using these simple models[4,12].

Recently, we have introduced the Persistent Embryo Method (PEM)[13] and applied it to the investigation of the nucleation of pure Ni and binary CuZr by all atom MD simulations. In this paper, we demonstrate that PEM is an efficient method by applying it to the Weeks-Chandler-Andersen (WCA) model[14], assuming the validity of CNT. As we describe further below, PEM can measure the nucleation rate in the low-density regime and does not make any geometrical assumption on the shape of nucleus. Moreover, PEM can obtain the unbiased configuration of the critical nucleus. We expect that PEM will become the method of choice as the optimal implementation of CNT for certain problems. Furthermore, given the current disagreements between nucleation rates as measured in experiments or calculated by theory, new techniques are crucial, in that they provide additional insights on the origins of such discrepancies.

This paper is organized as follows: in Sec. 2 we present the WCA model, including the description of Brownian dynamics and order parameter. In Sec. 3 we show the method and discuss the calculation of other necessary quantities, such as the chemical potential. Sec. 4 describes the persistent-embryo method (PEM) and Sec. 5 describes the determination of attachment rate. In Sec. 6, we derive the rate equation and present our predictions. In Sec. 7, we discuss our results and compare them with existing estimates obtained by other methods. A comparison with experimental results is also included. The conclusions are left for Sec. 8.

## 2. WCA model and simulation details

To relate our calculation to experiments[9–11], and also to compare with previous computational work[4–6], we performed Brownian dynamics (BD) simulations of N particles interacting with the WCA potential[14]:

$$U_{WCA}(r) = \begin{cases} 4\varepsilon\left[\left(\frac{\sigma}{r}\right)^{12} - \left(\frac{\sigma}{r}\right)^6 + \frac{1}{4}\right] & r \leq 2^{\frac{1}{6}}\sigma \\ 0 & r > 2^{\frac{1}{6}}\sigma \end{cases}$$

where $\sigma$ is the unit of length and $\varepsilon$ is the unit of energy. The WCA model is a softer version of the HS system[14], and it is possible to define a mapping from the WCA model to the HS model, as we discuss further below.

Brownian dynamics is the overdamped limit of Langevin dynamics. In Brownian dynamics, the equation of motion of particle $i$ is:

$$\frac{d\mathbf{r}_i}{dt} = \frac{1}{M\gamma}[-\nabla_i U + \mathbf{W}_i(t)] \quad (1)$$

where $\gamma$ is the friction coefficient with units of inverse time. $\mathbf{W}_i(t)$ is the stochastic force and $M$ is the mass of the particle, which is subject to a conservative force $-\nabla_i U$. The stochastic force is correlated according to the dissipation-fluctuation theorem $\langle \mathbf{W}_i(t)\mathbf{W}_j(t')\rangle = 6M\gamma k_B T \delta_{ij}\delta(t-t')$, where the $\delta$ is the Kronecker delta function and $k_B$ is the Boltzmann constant. The timestep of the simulation is $0.0004\,\tau$ ($\tau$ is the LJ time unit which equals $\sqrt{\frac{m\sigma^2}{\varepsilon}}$, where m is the unit of mass). Note the Brownian time $\tau_B$ is defined as $\tau_B = \frac{\sigma^2}{D}$, where $D$ is the diffusion coefficient on the Brownian motion determined by the Einstein relation $D = k_B T/M\gamma$. Following Kawasaki and Tanaka[5], we studied the WCA model at the reduced temperature $\beta\varepsilon = 40$, i.e. $T = 0.025\,\varepsilon/k_B$. For simplicity, both $M$ and $\gamma$ were set to unity. Therefore, $D = 0.025\,\frac{\sigma^2}{\tau}$ and $\tau_B = 40\tau$. All the simulations were performed using the GPU-accelerated LAMMPS code[15–17]. Some of the results were also tested using HOOMD-blue[18,19].

Identification of fluid-like and solid-like particles was accomplished by the use of the bond-orientational order parameter[20,21], which is described below for the sake of completeness. For each particle $i$, the bond-orientational order parameter is:

$$q_{6m}(i) = \frac{1}{N_b(i)} \times \sum_{j=1}^{N_b(i)} Y_{lm}(\vec{r}_{ij}) \qquad (2)$$

where $Y_{lm}$ are the spherical harmonics, $\vec{r}_{ij}$ is the position vector of particle j regards to particle $i$, and $N_b(i)$ is the number of nearest neighbors of particle $i$. The normalized bond-orientational order parameter is:

$$\tilde{q}_{6m}(i) = \frac{q_{6m}(i)}{[\sum_{m=-6}^{6}|q_{6m}(i)|^2]^{1/2}} \qquad (3)$$

A correlation between two particles $(i, j)$ is defined by:

$$S_{ij} = \sum_{m=-6}^{6} q_{6m}(i) q_{6m}^*(j) \qquad (4)$$

If $S_{ij}$ exceeds a threshold, then the two particles are considered bonded. To compare with a previous study[4], we chose $N_b(i)$ as the number of particles within a cutoff of $1.5\sigma$, and a threshold value for $S_{ij}$ of 0.7. We denote $\xi$ as the minimal number of connected neighbors for a particle to be considered as solid. In our simulation, if a given particle has 6 or more connected neighbors, it is considered as a solid-like particle, i.e., $\xi = 6$. This choice is somewhat arbitrary, and we will show later it leads to a significant uncertainty in the calculation of related physical quantities, as we explore the case of $\xi = 8$ and 9, and make an additional comparison with Filion *et al.*[4].

## 3. Chemical potential differences

The chemical potential difference between solid and fluid ($\Delta\mu$) is required for calculation of nucleation rates by CNT. To compute $\Delta\mu$, we make use of the Gibbs-Helmholtz integration:

$$\frac{\Delta\mu}{T_f} = \int_{T_m}^{T_f} \frac{\Delta H(T)}{T^2} dT \qquad (5)$$

where $\Delta H(T)$ is the enthalpy difference between the solid and fluid *under the target pressure*, $T_m$ is the melting temperature, and $T_f$ is final target temperature. The NPT ensemble was applied to calculate enthalpy (H). For a fixed density, we first ran a BD simulation at the target temperature ($\beta\varepsilon = 40$) to obtain the target pressure $P_f$ in the fluid for subsequent simulations. The target pressure in different densities was presented in Table 1, in the reduced form. To perform the constant pressure simulation for Brownian particles, we used the Berendsen barostat[22]. We ran *separate* simulations with 4000 particles in a solid face-centered cubic (FCC) phase and a fluid phase at the same temperature. We considered the FCC phase as it is well established that it has the lowest free energy among all putative crystalline phases in HS system. Other possibilities, such as body-centered cubic (BCC) and icosahedral phases, do not seem to play a role in the nucleation of monodisperse HS system[3]. As a cross-check, we ran simulations with a BCC embryo and found that it did not nucleate a crystal (see Appendix B).

To prepare the supercooled fluid phase, we first melted a crystal at a high temperature and cooled it down to the target temperature. The numerical integration was carried out using a chained trapezoidal rule with the interval of temperature $\Delta T = 0.0001\ \varepsilon/k_B$.

The melting temperature, as a function of fluid density, is the lower limit of Gibbs-Helmholtz integration. To determine the melting point, we ran BD simulation on a solid-fluid interface generated by melting half of the initial crystal (FCC) structure. During the simulation, the periodic boundary condition was applied, and we only allowed the length of the box in the direction perpendicular to the interface to change, resulting in constant interfacial area, constant normal pressure: $NP_xAT$ ensemble, with $x$ being the perpendicular direction. A typical configuration of $NP_xAT$ simulation is shown in Fig. 1. We monitored the length of the simulation cell $L_x$ along the x-direction to locate the phase transition at different temperatures: increasing of $L_x$ implies melting and vice versa.

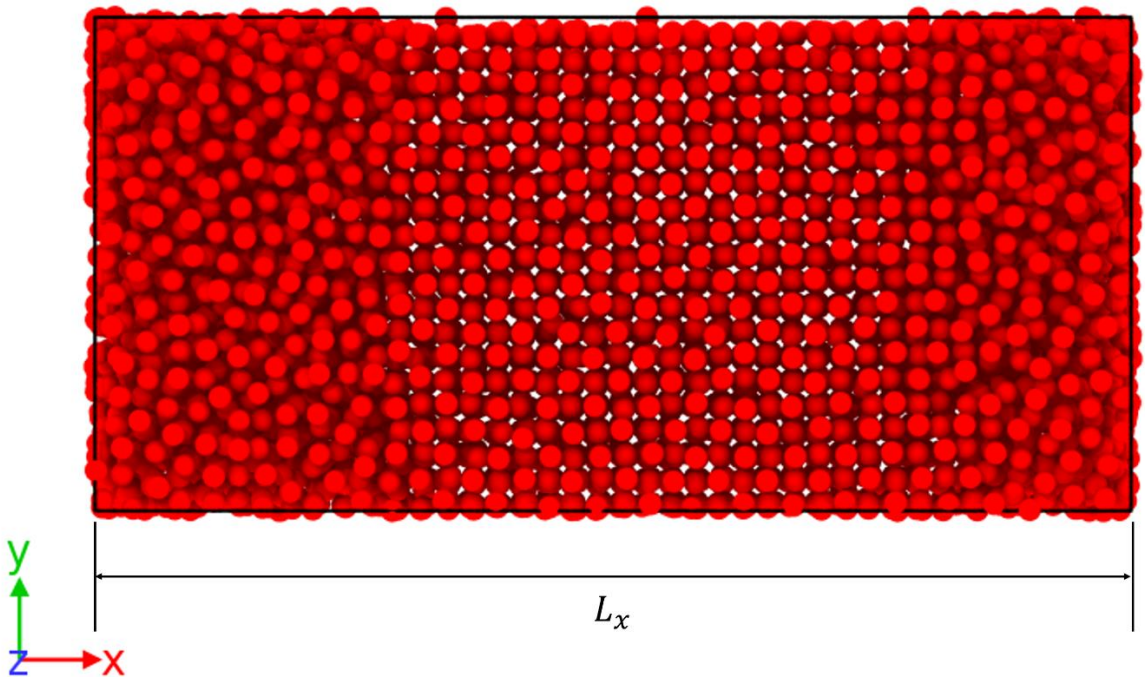

Fig. 1: Typical configuration of NPxAT simulation. The central part is the crystalline phase, while the remaining fluid. Periodic boundary conditions were applied during simulation. $L_x$ is the length of simulation box along x-direction.

A typical measurement of the melting point is exemplified in Fig. 2: At $T = 0.0319\ \varepsilon/k_B$, the simulation box expands, indicating that the solid is melting towards the fluid. At $T = 0.0317\ \varepsilon/k_B$, the simulation box shrinks, implying that the fluid side is crystalizing towards the solid. Only at $T = 0.0318\ \varepsilon/k_B$, the box size stays around $35.7\ \sigma$ and both phases coexist. Therefore, this is the melting temperature (within a $0.0001\ \varepsilon/k_B$ precision) at a fluid density $\rho\sigma^3 = 0.7525$. All the melting points thus obtained are shown in Fig. 2(b).

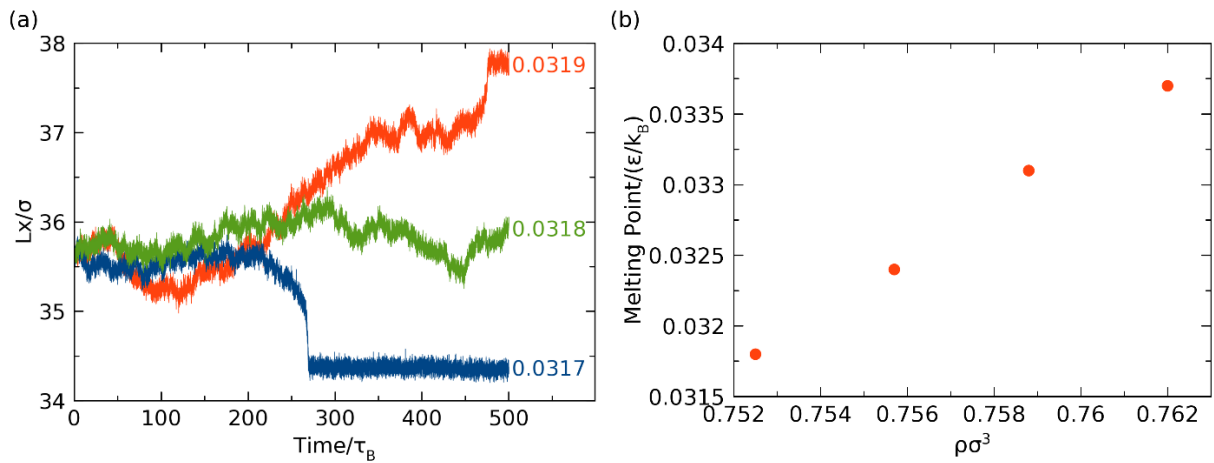

Fig. 2: (a) $L_x$ as a function of temperature at $T = 0.0317,\ 0.0318, 0.0319\ \epsilon/k_B$. (b) Melting point as a function of reduced density ($\rho\sigma^3$).

With the measured melting points and enthalpies, we compute the chemical potential differences by Equation 5. A summary of thermodynamic properties is given in Table 1. In Fig.3, we compared the reduced chemical potential differences with the results from Filion *et al.*[4], obtained by a thermodynamic integration of the equation of state[23]. Both methods provide consistent results.

Table 1. Thermodynamic properties calculated at various reduced densities, including the reduced pressure ($\beta p\sigma^3$), melting points ($k_B T_m/\varepsilon$) and reduced chemical potential differences ($\beta|\Delta\mu|$).

| $\rho\sigma^3$ | $\beta p\sigma^3$ | $k_B T_m/\varepsilon$ | $\beta|\Delta\mu|$ |
|---|---|---|---|
| 0.7525 | 11.34 | 0.0318 | 0.3298 |
| 0.7557 | 11.58 | 0.0324 | 0.3591 |
| 0.7588 | 11.83 | 0.0331 | 0.3925 |
| 0.7620 | 12.04 | 0.0337 | 0.4208 |

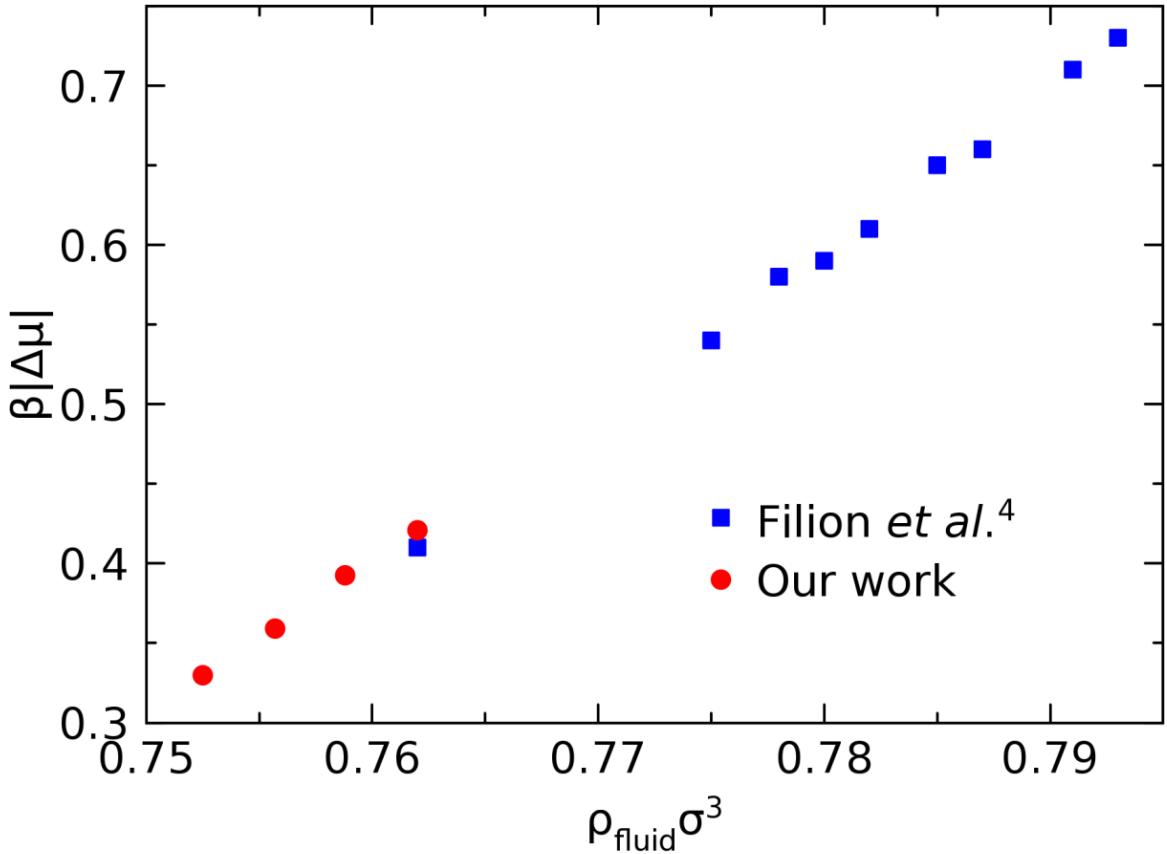

Fig. 3: The reduced chemical potential differences as a function of fluid density. The data from ref. 4 is also included in this figure.

## 4. The persistent embryo method (PEM)

Nucleation is described as a competition between the favorable fluid-to-crystal transition against the unfavorable cost to create the solid-fluid interface (SFI), leading to an excess free energy ($\Delta G$):

$$\Delta G(N) = N\Delta\mu + s\left(\frac{N}{\rho}\right)^{\frac{2}{3}}\gamma_i \qquad (6)$$

where $\Delta\mu(<0)$ is the chemical potential difference. $\gamma_i(>0)$ is the SFI free energy, and s is a geometric factor that accounts for the possible non-sphericity of the critical nucleus.

During PEM simulation, the number of total particles was set to 13500. The NVT ensemble was applied during PEM simulation. We first ran a BD simulation for 50,000 $\tau_B$ at $\rho\sigma^3 = 0.7525$ and we did not observe a single nucleation event. PEM[13] starts with the creation of a small embryo consisting of a crystal of $N_0$ atoms ($N_0 \ll N^*$), where $N^*$ is the size of critical nucleus. We inserted this embryo into the supercooled fluid and found that without any other modification, the embryo quickly dissolves back into the fluid. We therefore attached each of the *initial* embryo particles to a tunable spring while the other end of the spring is fixed to the embryo particle's equilibrium position, so that the embryo is prevented from shrinking and forced to grow. As the embryo grows, springs are gradually softened so that at a size $N_{sc}(<N^*)$ they are completely removed. This was accomplished by parameterizing spring constants according to:

$$k(N) = \begin{cases} k_0 \dfrac{N_{sc} - N}{N_{sc}} & if\ N < N_{sc} \\ 0 & otherwise \end{cases} \qquad (7)$$

where $N$ is the number of (solid-like) atoms within the nucleus. We ensured that the tunable parameter $N_{sc}$ is much smaller than critical nucleus size ($N^*$). Some "trial and error" experimentation is needed to determine the $N_0$ and $N_{sc}$ if an estimation of $N^*$ is unavailable.

To prepare the sample for our simulation, first we ran a BD simulation to melt the FCC crystal to a fluid phase. Then we selected an arbitrary particle as the center and inserted the crystalline embryo (mostly in FCC phase) into the fluid and deleted the fluid particles which are nearest neighbors to each particle of initial embryo. During this procedure, we added some crystalline particles into the system while removing the same number of fluid particles, so that the overall density remained constant.

Once the initial configuration was prepared, we started the relaxation of the entire system. First, we fixed the embryo particles and only heated the fluid particles to a relatively high temperature, and then annealed the fluid particles at high temperature. Finally, we quenched the fluid particles down to the target temperature ($\beta\epsilon = 40$). This procedure is designed to equilibrate the solid-fluid interface and repel any fluid particles that could be trapped inside the embryo. This preparation process allowed the system to reach thermal equilibrium for further PEM implementation.

We monitored the number of solid-like particles and updated the spring constant every 10,000 steps ($0.1\tau_B$), which we denoted as a loop. At the end of each loop, the size of the crystalline nucleus ($N$) was recalculated, and the spring constant was updated according to Equation 7. We terminated the simulation if the number of solid-like atoms grew too large (of the order of $5N_{sc}$), as the system was irreversibly crystallized.

The critical nucleus size ($N^*$) was computed by the following algorithm: as shown in Fig. 4, we firstly plotted the embryo size ($N$) versus time. According to the CNT, when the size of nucleus reaches $N^*$, $\Delta G$ reaches a maximum: The nucleus has equal probability to either dissolve or further grow. Therefore, the nucleus size will fluctuate around $N^*$ over a significant period of time, reflected as a plateau in the $N(t)$ times series. To guarantee that the selected plateaus correspond to the critical nucleus, we monitored the height and time width for each plateau. Occasionally, there are sporadic plateaus driven by thermal fluctuations, but those are relatively low and narrow. After a plateau is identified, the average is computed and assigned the critical value $N^*$. We launched many independent simulations (up to 60) to generate significant statistics. A collection of all plateaus is provided in Appendix A. For each plateau, we calculated its average height to cancel out the thermal fluctuation. The critical nucleus size $N^*$, as well as the standard deviation of the measurement, is then determined by averaging the heights of all the plateaus.

Fig. 4 shows a typical PEM simulation, where we show the embryo size versus time at $\rho\sigma^3 = 0.7525$. $\xi = 6$ was applied in this case. There are several critical plateaus before the nucleus irreversibly grows. The actual critical nucleus including all the atoms shows a clear anisotropic shape with obvious crystalline facets.

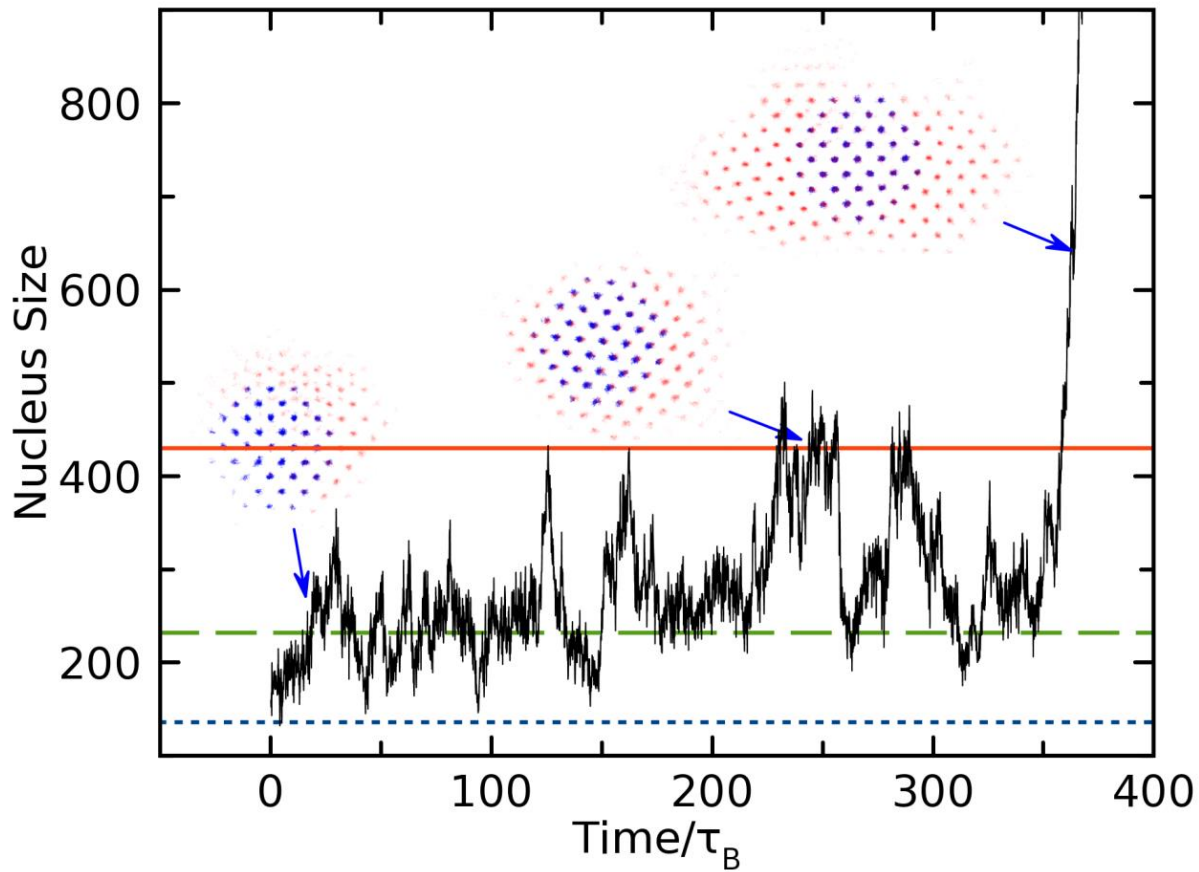

Fig. 4: Nucleus size versus time in the PEM simulation. The dot line (blue) shows the number of atoms in the embryo($N_0$), and dash line (green) indicates the threshold to remove the springs($N_{sc}$). The solid line corresponds to the critical nucleus size ($N^*$) which is determined statistically from 8 plateaus (see Appendix A for the collection). The three insets show the time-averaged shape of pre-critical, critical and post critical nuclei at the corresponding time (arrow). The blue dots indicate the embryo and red ones are particles attached to the embryo.

We considered at least 8 plateaus at every density. The averaged critical nucleus size $N^*$ and the standard deviation are summarized in Table 2. A comparison between our result and ref. 8 is given in Fig. 5. Overall, our results of $\xi = 6$ are in better agreement with those obtained by Forward Flux Sampling (FFS)[8] than those from the seeding method[8]. The deviation between PEM and the seeding method may originate from the geometrical assumption of critical nucleus in seeding method. Another comparison between PEM and the seeding method can be found in the supplementary material of ref. 13. To make a comparison, both results from $\xi = 6, 8$ and 9 are included in Fig. 5.

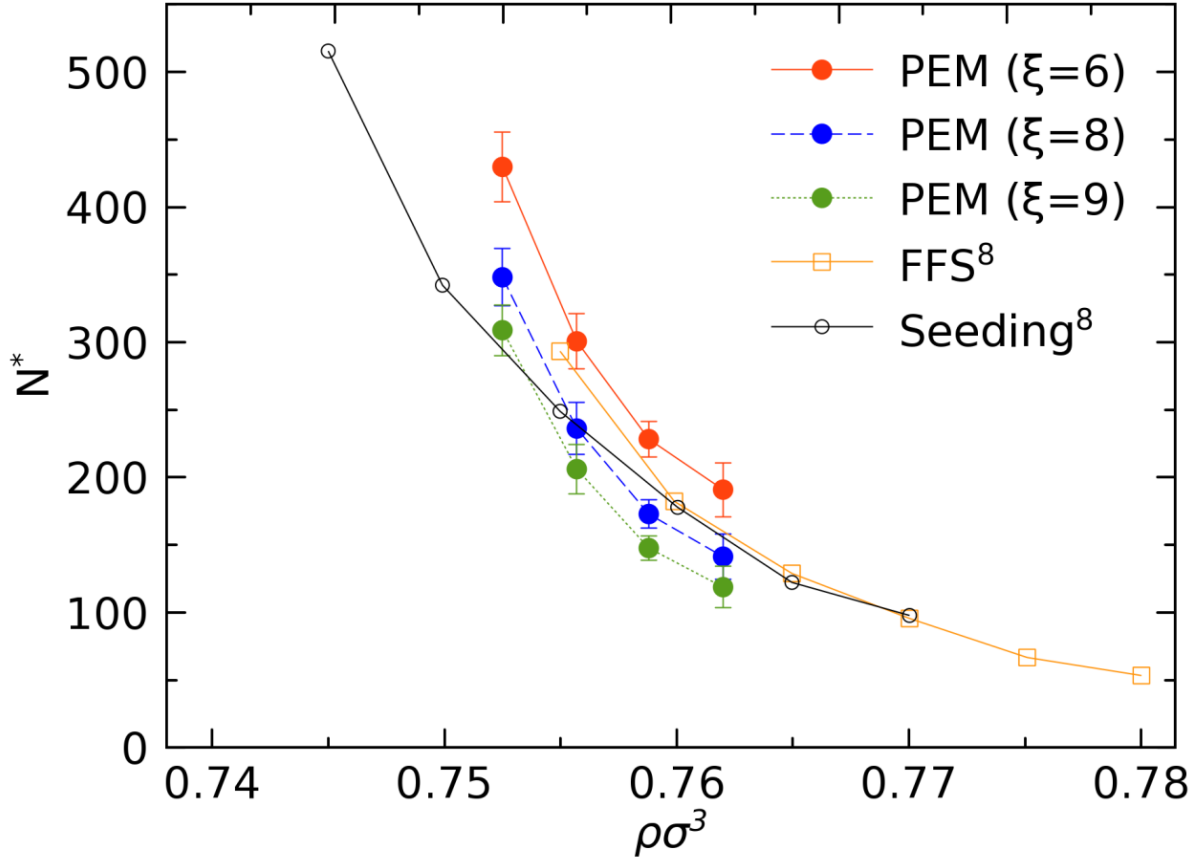

Fig. 5: Critical nucleus size ($N^*$) as a function of reduced (fluid) density ($\rho\sigma^3$). Both $\xi = 6$ (red solid line), $\xi = 8$ (blue dash line) and $\xi = 9$ (green dot line) are considered. Previous results are given by either Forward Flux Sampling (FFS) or the seeding method from ref. 8, using $\xi = 9$.

Table 2: Estimates for the nucleus size ($N^*$) and its standard derivation ($\Delta N^*$), attachment rate ($f^+\tau_B$). $N^*$ and $\Delta N^*$ are calculated in terms of $\xi = 6, 8$ and 9, as shown explicitly.

| $\rho\sigma^3$ | $N^* \pm \Delta N^*$ ($\xi = 6$) | $N^* \pm \Delta N^*$ ($\xi = 8$) | $N^* \pm \Delta N^*$ ($\xi = 9$) | $f^+\tau_B$ |
|---|---|---|---|---|
| 0.7525 | 429.76 ± 26.4 | 348.06 ± 21.1 | 308.89 ± 18.8 | 490.7 |
| 0.7557 | 300.67 ± 20.3 | 236.13 ± 19.3 | 206.09 ± 18.1 | 486.9 |
| 0.7588 | 228.30 ± 14.0 | 172.85 ± 10.6 | 147.65 ± 9.1 | 436.4 |
| 0.762 | 190.91 ± 19.9 | 141.31 ± 16.8 | 118.73 ± 15.3 | 437.8 |

## 5. Attachment rate

The attachment rate at the critical nucleus was calculated by the method first introduced by Auer and Frenkel[24]. It considers the size change of the critical nucleus, based on the iso-configurational ensemble[25], which is given by the formula:

$$f^+ = \frac{\langle |N(t) - N^*|^2 \rangle}{2t} \qquad (8)$$

To determine the attachment rate, we selected ta snapshot from PEM simulations whose nucleus size is exactly $N^*$ as the initial configuration. We then performed 50 independent runs starting from this configuration. Different trajectories were reached in these runs because of the random noise introduced by reinitializing the random force term $\boldsymbol{W(t)}$ in the Langevin equation for each particle. (The springs are absent.)

Fig. 6 shows a determination of attachment rates at $\rho\sigma^3 = 0.7525$. From the upper panel we notice that roughly half of the runs ended up melting back into the fluid, while the other half grew irreversibly, as expected. Further confirmation is obtained by plotting the ensemble average $\Delta N^*(t) = N(t) - N^*$. According to the CNT, $\frac{d\langle \Delta N^*(t) \rangle}{dt} = 0$ at $N^*$, which is clearly shown in Fig. 6. Therefore, we can measure the attachment rate by fitting the slope of $|\Delta N^*(t)|^2$ to Equation 8. The obtained attachment rates for different densities are summarized in Table 2.

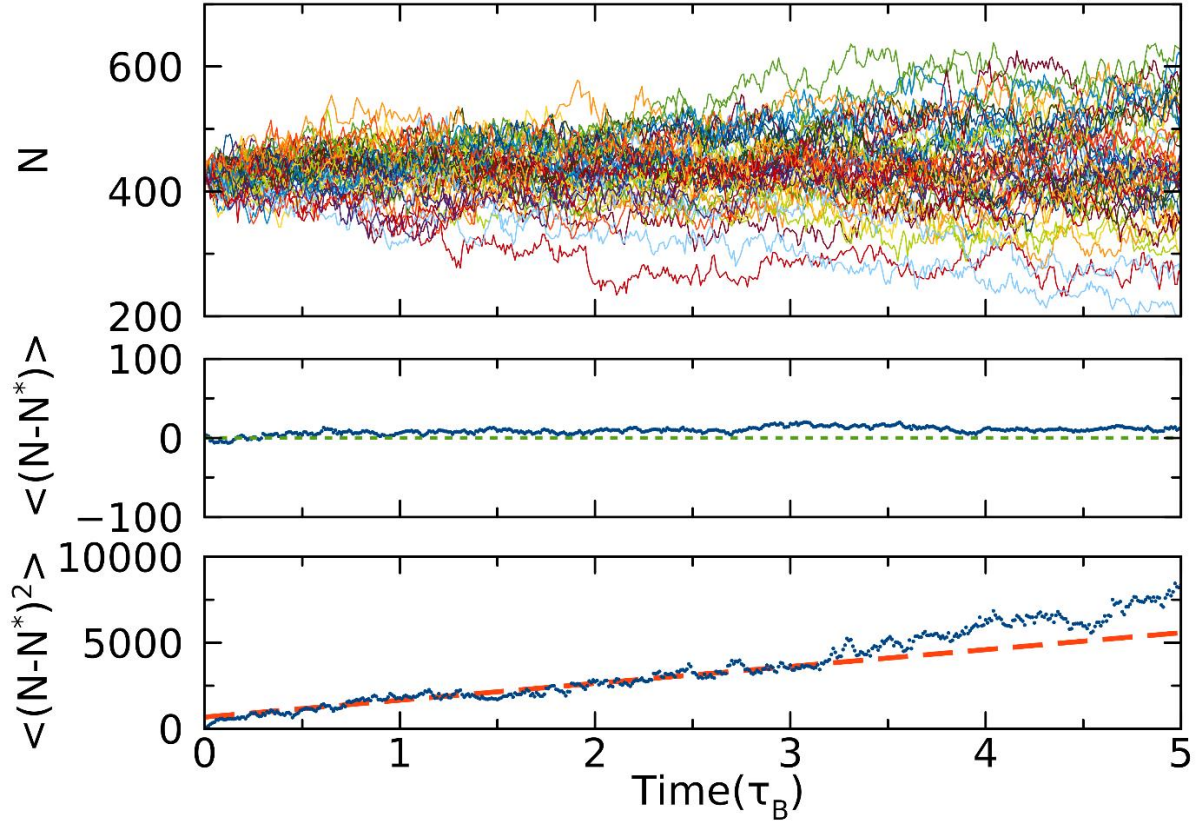

Fig. 6: The upper panel shows the nucleus size versus time for the isoconfigurational ensemble including 50 runs. The middle panel shows the ensemble average of $\Delta N^*(t) = N(t) - N^*$. The dot line (green) serves as a guide to the eye for the zero point. The bottom panel shows the ensemble average of $|\Delta N^*(t)|^2 = |N(t) - N^*|^2$. The dashed line (red) indicates the linear fitting to the range from $0.5\tau_B$ to $3\tau_B$ to derive the attachment rate. During $0.5\tau_B$ to $3\tau_B$, $\Delta N^*(t)$ remains to be almost zero.

## 6. Nucleation rate

The nucleation rate J is given by:

$$J = \kappa \exp(-\frac{\Delta G^*}{k_B T}) \qquad (9)$$

where $\kappa$ is a kinetic prefactor. $\Delta G^*$ depends on the driving force $|\Delta \mu|$ and the critical nucleus size $N^*$ by[13]:

$$\Delta G^* = \frac{1}{2} N^* |\Delta \mu| \qquad (10)$$

The explicit formula for the nucleation rate is[26]:

$$J = \rho_L f^+ \sqrt{\frac{|\Delta \mu|}{6\pi k_B T N^*}} \exp\left(-\frac{|\Delta \mu| N^*}{2 k_B T}\right) \qquad (11)$$

Nucleation rates were computed from Equation 11 and are summarized in Table 3. With nucleation rates of the order of $10^{-15}$ or even smaller, it is not surprising that no nucleation events were observed by brute-force simulation alone.

One should notice that the rate $J$ depends on $N^*$ exponentially, so a small uncertainty in $N^*$ will result in a significant variation in $J$. Our calculation in terms of $\xi = 6, 8$ and 9 reconfirms this statement. It is worth mentioning that PEM rigorously identifies the configuration of critical nucleus. The error only enters in the identification of the number of solid-like particles. Further discussion about the order parameter can be found in ref. 27–29.

Table 3: Mapped volume fraction ($\phi_{eff}$, see below for its definition) and nucleation rate ($J\sigma^5/D$) as a function of reduced density ($\rho\sigma^3$). Nucleation rate was calculated in the case of $\xi = 6, 8$ and 9.

| $\rho\sigma^3$ | $\phi_{eff}$ | $J\sigma^5/D$ ($\xi = 6$) | $J\sigma^5/D$ ($\xi = 8$) | $J\sigma^5/D$ ($\xi = 9$) |
|---|---|---|---|---|
| 0.7525 | 0.5200 | $3.934 \times 10^{-31}$ | $3.103 \times 10^{-25}$ | $2.104 \times 10^{-22}$ |
| 0.7557 | 0.5222 | $1.051 \times 10^{-23}$ | $1.280 \times 10^{-18}$ | $1.917 \times 10^{-16}$ |
| 0.7588 | 0.5243 | $1.097 \times 10^{-19}$ | $6.715 \times 10^{-15}$ | $1.022 \times 10^{-12}$ |
| 0.762 | 0.5265 | $1.294 \times 10^{-17}$ | $5.120 \times 10^{-13}$ | $2.185 \times 10^{-11}$ |

## 7. Discussion

Computed nucleation rates are compared to other available estimates in Fig. 7. At a density of $\rho\sigma^3 = 0.762$, the nucleation rates given by Filion *et al.*[4] using US and FFS are in agreement (within error bar) to our PEM estimates. At the smaller density, PEM gives a slightly lower nucleation rate than FFS from ref. 6. One should notice that the PEM calculations used the NVT ensemble, while the US and FFS in ref. 4 operated in the NPT ensemble. As showed in Table 3, the critical nucleus size at the density of $\rho\sigma^3 = 0.762$ is $N^* = 190.91$, while the total number of system is 13500. When the nucleus reaches the critical size ($N^*$), it is only 1% of the entire system. In this case, the fluctuations that differentiate the NVT and NPT are virtually negligible[5]. In NVT simulations, there is a feedback mechanism in that a growing crystalline nucleus has a higher density and thus reduces the overall pressure, which prevents it from growing any further. Thus, the NVT simulation actually evaluates the lower limit of the nucleation rates, as compared to the NPT case. The influence of order parameter is explicitly shown in the Fig. 5, and we will discuss it below. Besides, there is a recent paper[30] which reveals that although FFS is not sensitive to order parameter, the conventional FFS sometimes still underestimates the nucleation rate by several orders of magnitude.

Furthermore, we also measured the nucleation rate at large densities where the nucleation can be directly accessed by brute-force Brownian dynamics simulations. Mean-first passage time[31] are measured from 20 brute-force simulations. Computed nucleation rate is also shown in Fig. 7, which is in good agreement with existing estimates.

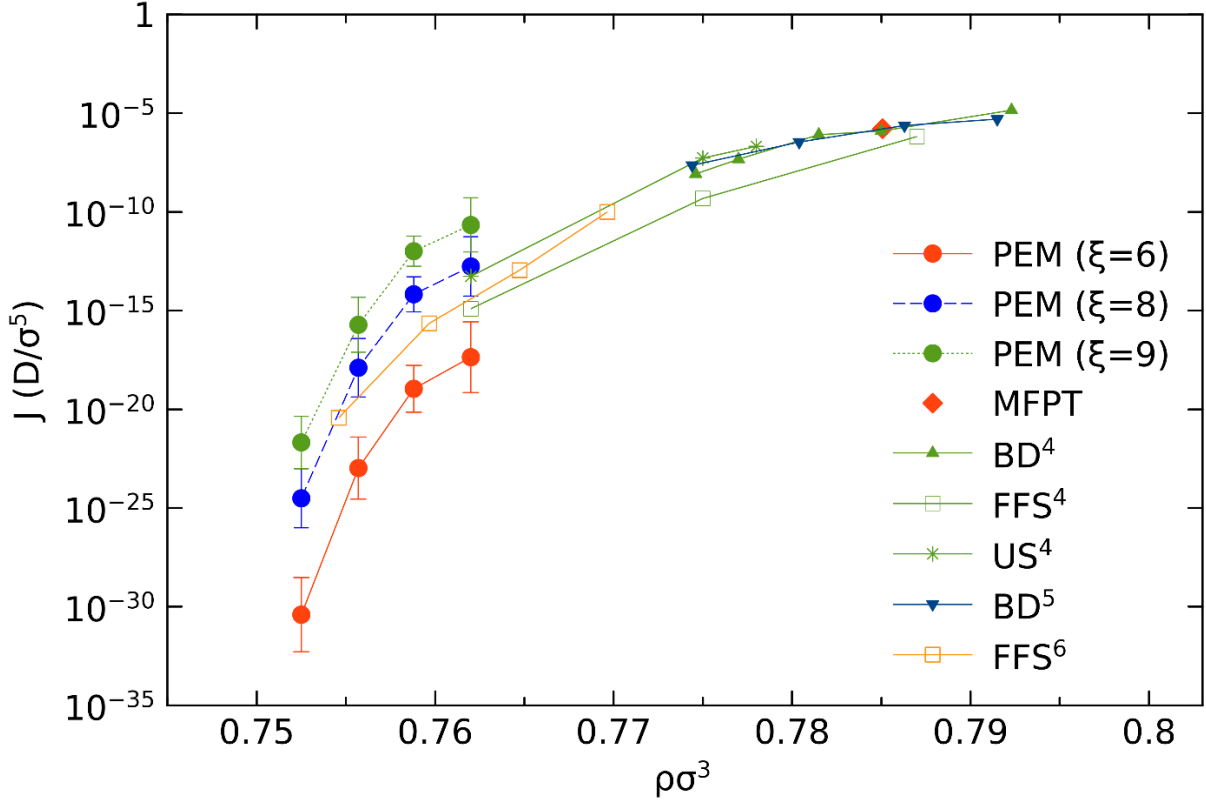

Fig. 7: The nucleation rate as a function of fluid density, in the case of $\xi = 6, 8$ and $9$. The results from previous work on WCA particles from ref. 4–6 are also included.

To translate the computed nucleation rates of WCA model with HS models, we map the density $\rho$ to the volume fraction $\phi$ by:

$$\phi = \left(\frac{\pi}{6}\right)\rho d^3 \quad (12)$$

where $d$ is the effective diameter of WCA particles. Here we follow the same procedure as described by Filion et al.[4] to map the freezing density $\rho_f \sigma^3 \simeq 0.712$ of WCA particles to the freezing volume fraction $\phi_F^{HS}$ of hard spheres. Even for the ideal HS model, there is some uncertainty about the exact value of the freezing volume fraction, which is only accurately determined within a range of $0.491 < \phi_F^{HS} < 0.494$[23,32,33]. In our study, we choose $\phi_F^{HS} \simeq 0.492$, as this seems to be the most precise value[23]. Thus, density and volume fraction are related by the effective diameter $d \simeq 1.097\sigma$. The mapped volume fraction is enclosed in Table 3. Note while the uncertainty in $\phi_F^{HS}$ appears to be quite small, it may result in a very large dispersion when mapping the number density to the volume fraction. In Fig. 8, the nucleation rates are shown as a function of volume fraction and compared to both experimental and other simulation results. The horizontal error bar indicates the lower and upper limit for volume fractions. Given the logarithmic nature of the plot, slight differences in the mapping result in large shifts of the nucleation rates. Another crucial factor is the order parameter, as indicated in Fig. 5, Fig. 7 and Fig. 8. A small change in $\xi$ can lead to a significant change in critical nucleus size ($N^*$) and consequently, the nucleation rate ($J$). But the advantage of PEM is that it can get the exact configuration of the critical nucleus, while the error is introduced in the counting of solid-like particles. Considering that the $J$ relies on $N^*$ exponentially, a small difference of $N^*$ can enhance a huge jump of $J$. Any method, as long as it relies on CNT, suffers from this problem. On the other hand, the bond-orientational order parameter is not the only representation of the solid particles within the nucleus. An alternative option is the $(n, v)$ parameter which incorporates a kinetic theory to the descriptor of nucleus and can be used to describe the dynamics of sub-critical nucleus[34,35].

Despite all these uncertainties, our WCA simulations are in clear agreement with previous estimates[2,4–6]. However, all the computed nucleation rates, including current PEM results, deviate from the experimental measurements[9–11]. As pointed out in ref. 4, this discrepancy is not related to inaccuracies of the theoretical calculation but rather, reflects a true deficiency of HS and related models to capture the physics of these experimental systems. Our calculations support this conclusion.

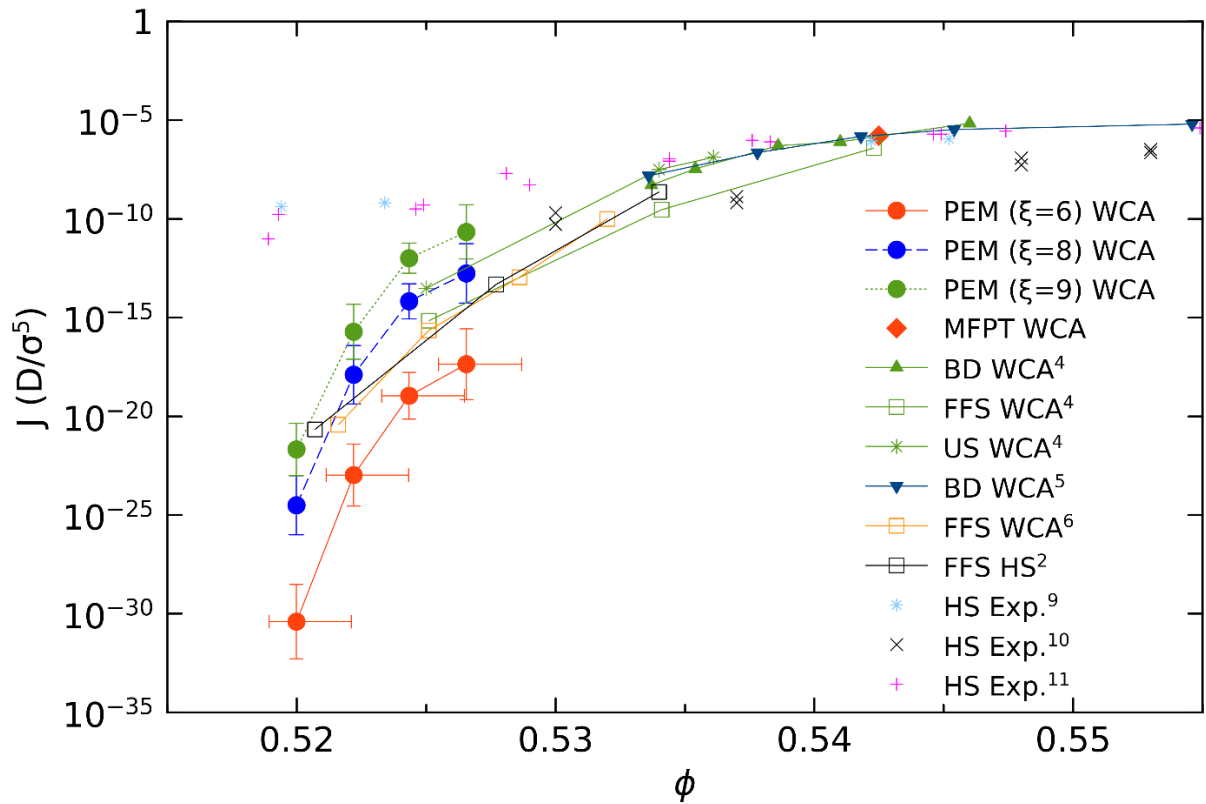

Fig. 8: Nucleation rates as a function of mapped volume fraction, for $\xi = 6$ (red solid line), $\xi = 8$ (blue dash line) and $\xi = 9$ (green dot line). Simulation resulting from WCA models[4–6], hard spheres[2] and experiments[9–11] are included in the figure. The horizontal error bar when $\xi = 8$ and 9 is the same with the case when $\xi = 6$.

## 8. Conclusions

We have measured the chemical potential differences between undercooled WCA fluids and solids by Gibbs-Helmholtz integration. We implemented the persistent-embryo method (PEM) to evaluate the nucleation rates in the density of $\rho\sigma^3 = 0.7525, 0.7557, 0.7588, 0.762$. The rates given by PEM are consistent with the ones available from umbrella sampling (US) and forward flux sampling (FFS). Our method-PEM is an efficient method to evaluate nucleation rates and provides a unique characterization of not only the critical, but the dynamics of the pre-critical and post-critical nucleus. Additionally, estimates obtained within WCA models are in agreement with hard spheres. Our results provide more evidence that the discrepancy between simulation and experiments will require more sophisticated models.

## Conflicts of interest

There are no conflicts to declare.

## Acknowledgements

Shang Ren would like to thank Xun Zha for help on HOOMD-blue. Work at Ames Laboratory was supported by the US Department of Energy, Basic Energy Sciences, Materials Science and Engineering Division, under Contract No. DE-AC02-07CH11358, including a grant of computer time at the National Energy Research Supercomputing Center (NERSC) in Berkeley, CA. K.M.H. acknowledges support from USTC Qian-Ren B (1000-Talents Program B) fund. The Laboratory Directed Research and Development (LDRD) program of Ames Laboratory supported the use of GPU-accelerated computing.

# Appendix A: Collection of plateaus

The collection of plateaus is presented in the Fig. A1.

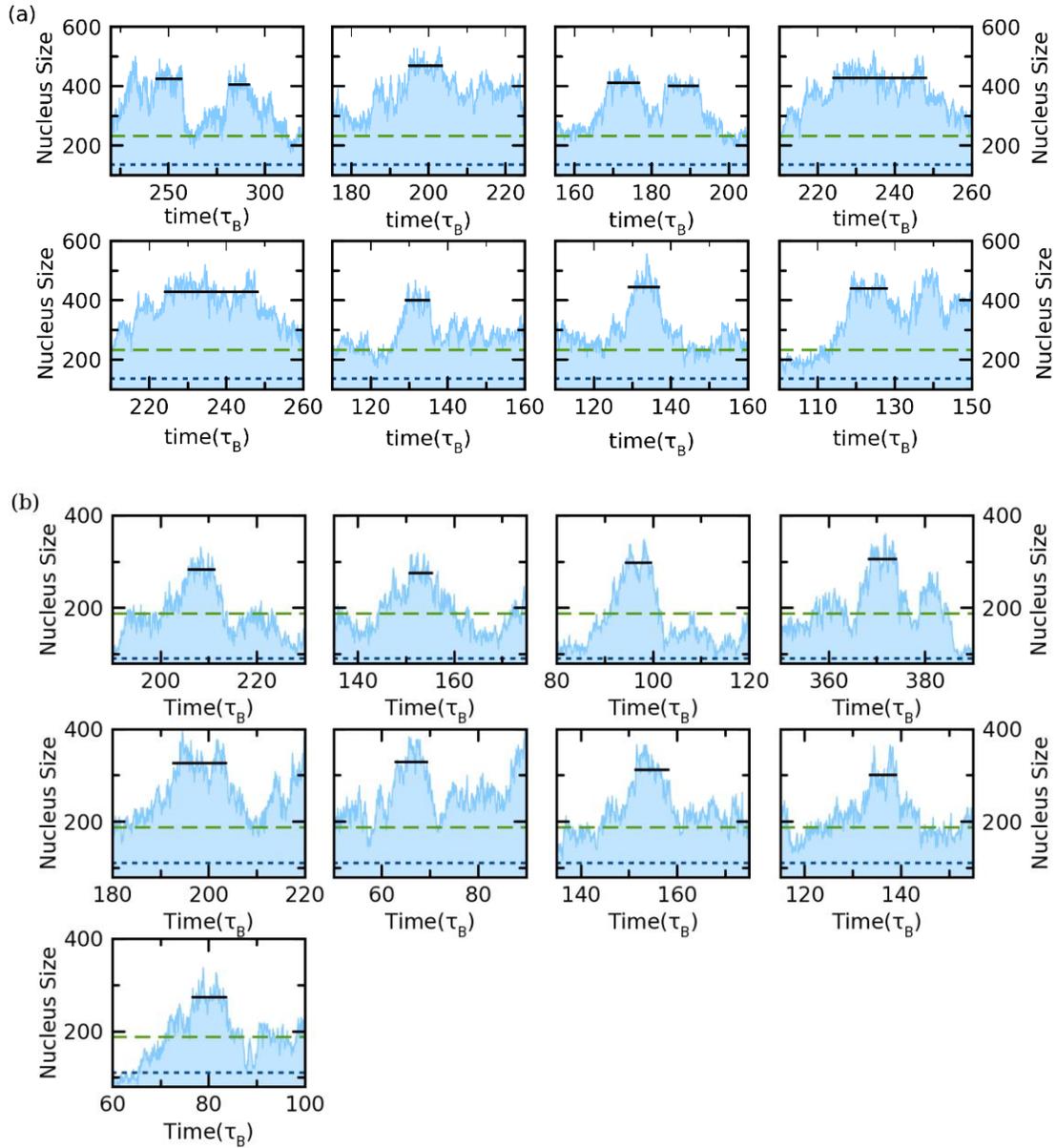

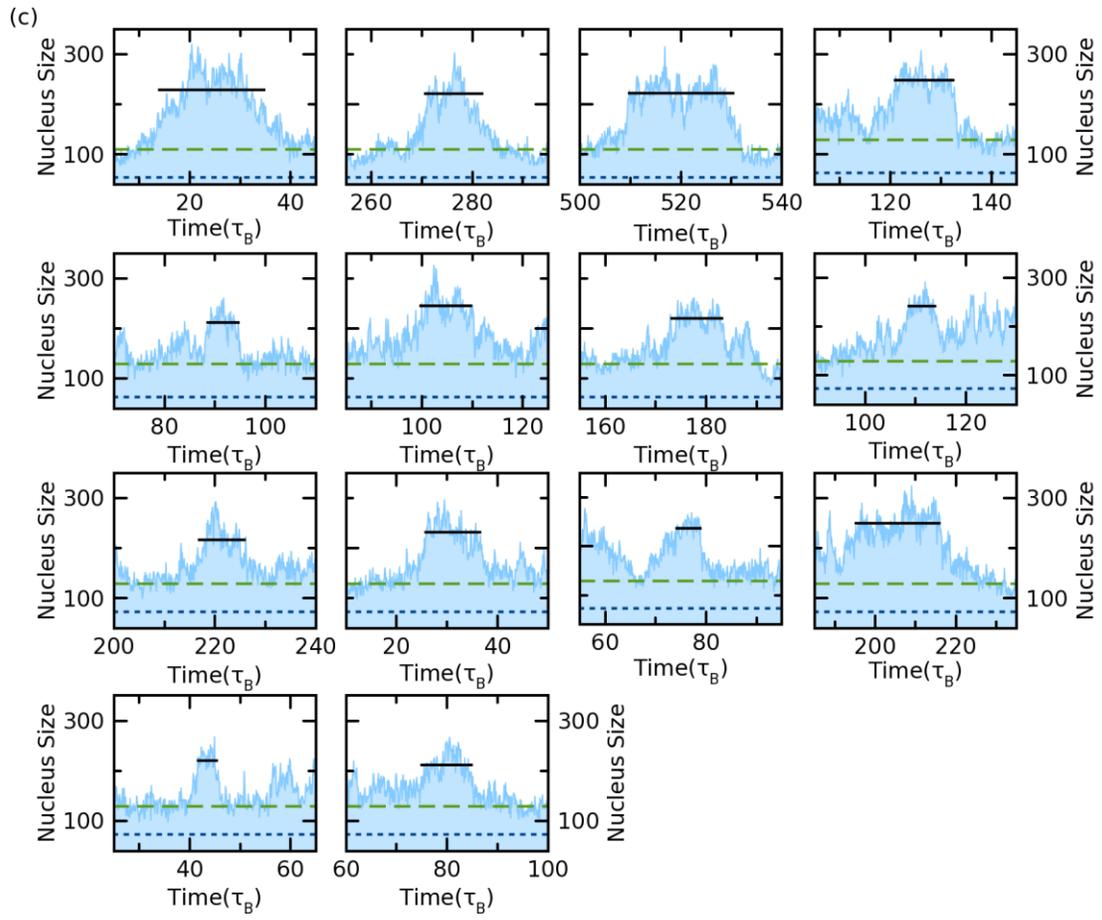

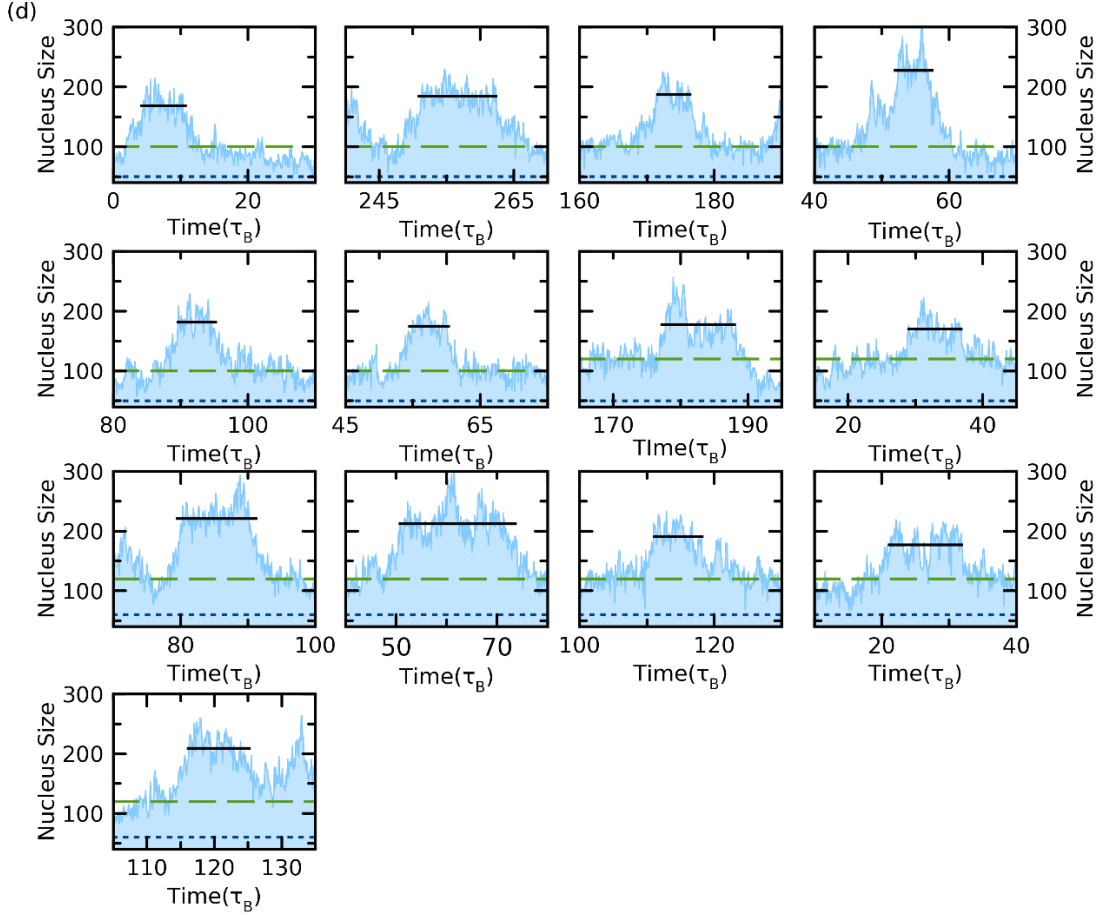

Fig. A1: The plateaus to determine the critical nucleus size in different reduced densities ($\rho\sigma^3$): (a) 0.7525, (b) 0.7557, (c) 0.7588, (d) 0.762. During the PEM simulation, different $N_0$, which is the size of embryo, and $N_{sc}$, which is the threshold where the spin is removed, were applied and indicated by red dot line and green dash line, respectively. The black solid line shows the averaged height of each plateau.

## Appendix B: The crystalline and geometrical structure of the embryo

To investigate the effect of geometry of embryo, we launched 36 independent PEM simulations in the condition that the embryo is cubic, and 12 plateaus were collected among them. By the same procedure, we find that the critical nucleus size ($N^*$) for $\rho\sigma^3 = 0.7525$ is $303.44 \pm 15.4$, while in the spherical embryo case the $N^* = 300.67 \pm 20.3$. The two ranges have huge overlaps which indicts that the geometry of embryo is not a critical issue in the PEM. A collection of plateaus is enclosed in the Fig. A2.

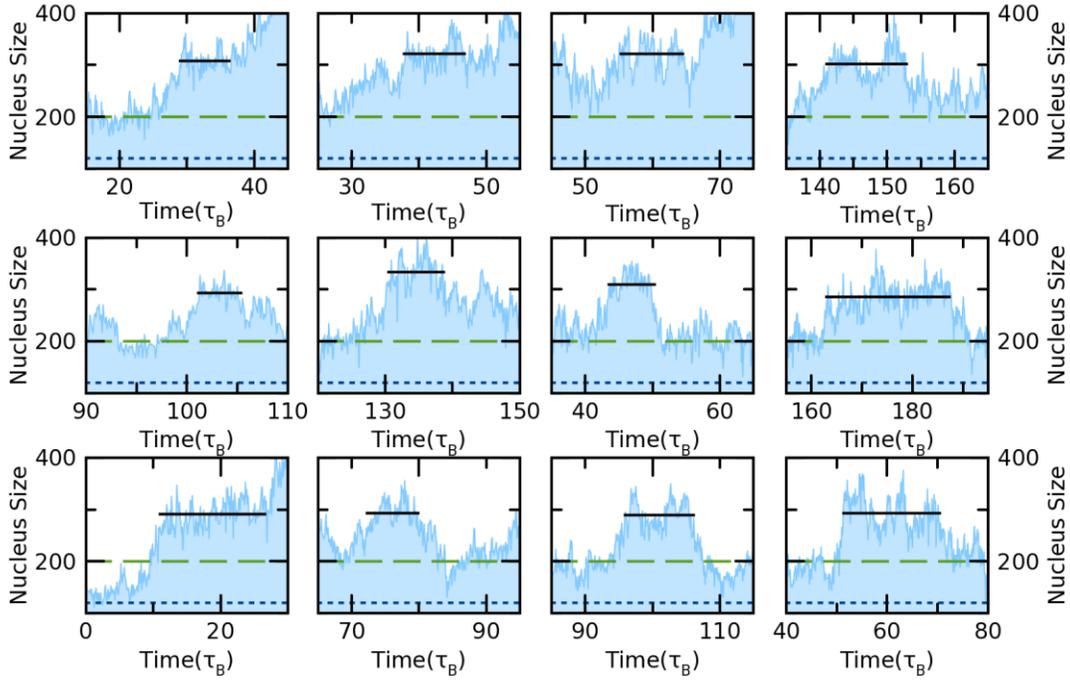

Fig. A2: The plateaus to determine the critical nucleus size in the case of cubic embryo. The (reduced) density is $\rho\sigma^3 = 0.7557$. During the PEM simulation, different $N_0$, which is the size of embryo, and $N_{sc}$, which is the threshold where the spin is removed, were applied and indicated by red dot line and green dash line, respectively. The black solid line shows the averaged height of each plateau.

Besides the geometry, another crucial factor is the structure of embryo, e.g. how does a BCC embryo work in the PEM? This question has been discussed by Auer and Frenkel[3], in the year of 2001. It is claimed that the BCC and icosahedral do not play a role in the nucleation process of HS system. To test this, we launched 12 independent runs to justify the effect of BCC. The result shows that the BCC embryo cannot grow while in the same time period the FCC embryo can lead the crystallization, as shown in Fig. A3. Therefore, the BCC is not dominant if the nucleus is relatively large and we excluded the nucleation channel via BCC in our simulation.

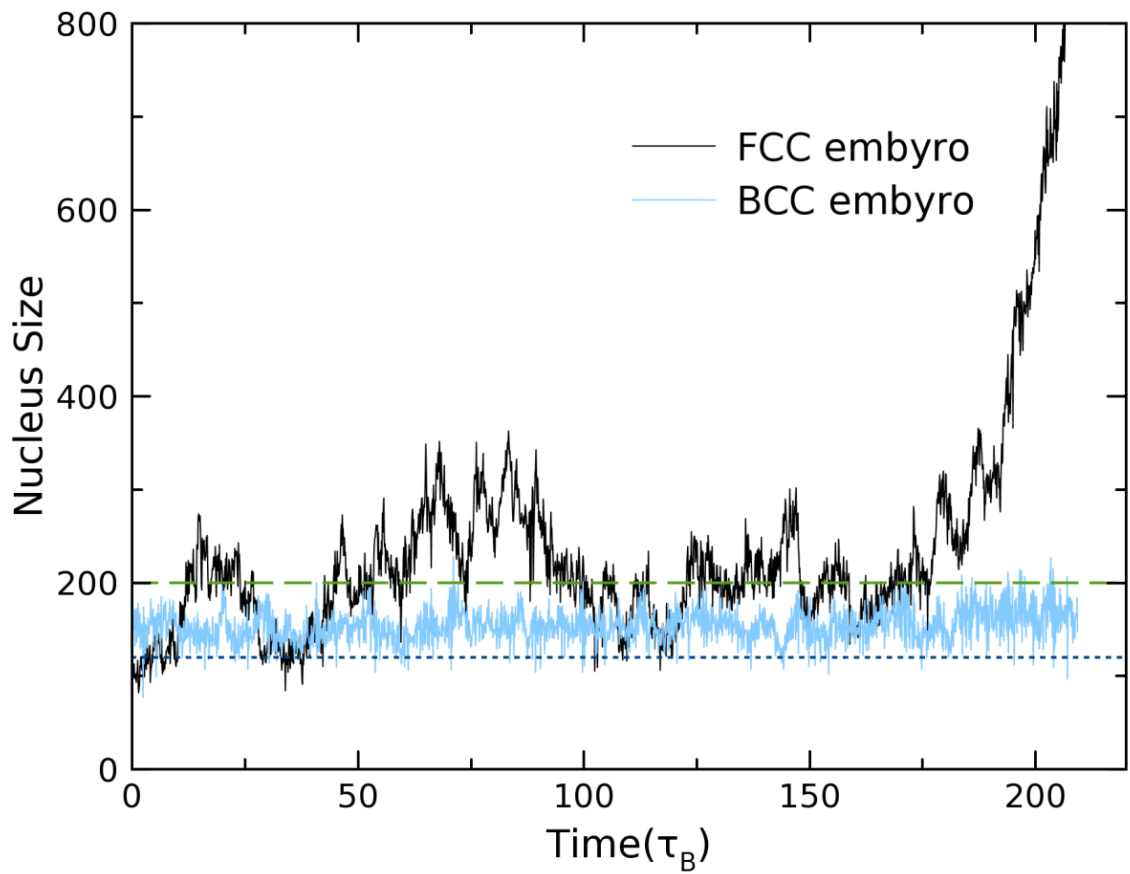

Fig. A3: (color online) A comparison of simulations starting with FCC embryo and BCC embryo. The black line corresponds to the FCC embryo, showing that the FCC embryo eventually grew. The blue line corresponds to the BCC embryo. Within simulation time, the BCC embryo did not grow.